\documentclass[preprint,superscriptaddress,showkeys,showpacs,
tightenlines,fleqn,nofootinbib,prd,byrevtex]{revtex4} 
\usepackage{graphicx}
\begin{document}      
\preprint{YITP-07-52}
\preprint{PNU-NTG-10/2007}
\preprint{INHA-NTG-02/2008}
\title{Electromagnetic form factors of the pion and 
kaon \\from the instanton vacuum}
\author{Seung-il Nam}
\email[E-mail:]{sinam@yukawa.kyoto-u.ac.jp}
\affiliation{Yukawa Institute for Theoretical Physics (YITP), 
Kyoto University, Kyoto 606-8502, Japan} 
\author{Hyun-Chul Kim}
\email[E-mail:]{hchkim@inha.ac.kr}
\affiliation{Department of Physics, Inha University, Incheon 402-751,
Republic of Korea}
\date{February 2008}
\begin{abstract}   
We investigate the pion and kaon ($\pi^+$, $K^+$, $K^0$)
electromagnetic form factors in the space-like region:  
$Q^2\lesssim1$ GeV, based on the gauged low-energy effective chiral
action from the instanton vacuum in the large $N_c$ limit.  Explicit
flavor SU(3) symmetry breaking is taken into account.  The
nonlocal contributions turn out to be crucial to reproduce the
experimental data.  While the pion electromagnetic form factor is in
good agreement with the data, the kaon one seems underestimated.  We
also calculate the electromagnetic charge radii for the pion and kaon:
$\langle r^2\rangle_{\pi^+}=0.455\,\mathrm{fm}^2$, $\langle
r^2\rangle_{K^+}=0.534\, \mathrm{fm}^2$ and $\langle 
r^2\rangle_{K^0}=-0.060\,\mathrm{fm}^2$ without any 
adjustable free parameter except for the average instanton size and
inter-instanton distance, and they are compatible with the
experimental data.  The low-energy constant $L_9$ in the large $N_c$
limit is estimated to be $8.42\times10^{-3}$ from the pion charge
radius.    
\end{abstract} 

\maketitle
\section{Introduction}
Understanding the structure of the pion has been one of the most
important issues in quantum chromodynamics (QCD), since it is the
lightest hadron, so that it is identified as the Goldstone boson
arising from spontaneous breaking of chiral symmetry.  In particular,
the electromagnetic (EM) form factor of the pion reveals its 
internal structure in terms of quark and gluon degrees of freedom.
Theoretically, it can be described at high energy space-like 
region of the momentum transfer by the factorization
theorem~\cite{Efremov:1979qk,Lepage:1979zb,Lepage:1980fj, 
Stefanis:1998dg,Stefanis:2000vd}, {\it i.e.} it can 
be explained by separating the hard perturbative amplitude from the
soft-part pion wave function which contains nonperturbative
information of QCD. At very small momentum transfers, chiral
perturbation theory ($\chi$PT) provides a good framework to explain
the pion EM form factor~\cite{Gasser:1983yg}.  A recent work has
calculated the pion and kaon form factors to next-to-next-to leading
order in $\chi$PT~\cite{Bijnens:2002hp} in the range of the momentum
transfer $-0.25\le q^2 \le 0\,{\rm GeV}^2$.  Moreover, the pion
EM form factor has been studied also in 
lattice QCD~\cite{Bonnet:2004fr,Capitani:2005ce,Brommel:2006ww} with
various different methods used.  Experimentally, there has
been a great deal of measurements of the pion EM form
factor~\cite{Dally:1977vt,Bebek:1974iz, Bebek:1974ww, Bebek:1977pe,
Molzon:1978py, Dally:1980dj,Dally:1982zk,
Amendolia:1986wj,Amendolia:1986ui,Liesenfeld:1999mv,Volmer:2000ek, 
Horn:2006tm, Tadevosyan:2007yd}.  

In the intermediate space-like region of the momentum transfer, one
has to use model approaches in order to investigate the pion EM form
factor, since neither perturbative QCD nor $\chi$PT can be applied to
this region.  Actually, there has been a great amount of theoretical
works to explain the pion EM form factor, various theoretical models
being employed: For example, the vector dominance
model~\cite{Klingl:1996by}, the Nambu-Jona-Lasinio (NJL) 
model~\cite{Ito:1991pv,Roberts:1994hh,Plant:1997jr}, Bethe-Salpeter
amplitudes~\cite{Maris:1999bh, Maris:2000sk}, Instanton
approach~\cite{Forkel:1994pf,Dorokhov:1991nj,Dorokhov:2003sc,
Dorokhov:2004ze,Faccioli:2002jd}, Kroll-Lee-Zumino
model~\cite{Dominguez:2007dm} and so on.   

In the present work, we want to investigate the pion and kaon EM form
factors in the space-like momentum transfer region ($0\le
Q^2\le$ GeV), 
based on the gauged low-energy effective chiral action (E$\chi$A)
from the instanton vacuum~\cite{Diakonov:1985eg}.  In order to
describe the kaon EM form factor, we need to consider SU(3) symmetry
breaking explicitly.  Thus, we employ the extended effective chiral
action developed in Refs.~\cite{Musakhanov:1998wp,Musakhanov:2001pc,  
  Musakhanov:2002vu,Musakhanov:2002xa} in which the effects of the
current quark mass have been explicitly considered in the instanton
vacuum.  Since the instanton vacuum realizes $\chi$SB naturally via 
quark zero modes, it may provide a good framework to study the EM form
factor of the pion and kaon, {\it i.e.} of the pseudo-Goldstone bosons.
Another virtue comes from the fact that there are only two parameters
in this approach, namely the average instanton size $\bar \rho\approx
\frac{1}{3}\, \mathrm{fm}$ and average inter-instanton distance $\bar
R\approx 1\, \mathrm{fm}$.  The normalization point of this approach
is determined by the average size of instantons and is approximately   
equal to $\rho^{-1}\approx 0.6\, \mathrm{GeV}$.  The values of
the $\bar \rho$ and $\bar R$ were estimated many years ago
phenomenologically in Ref.~\cite{Shuryak:1981ff} as well as
theoretically in
Ref.~\cite{Diakonov:1983hh,Diakonov:2002fq,Schafer:1996wv}.  
Furthermore, it was confirmed by various lattice simulations of
the QCD vacuum \cite{Chu:vi,Negele:1998ev,DeGrand:2001tm}.  Also
lattice calculations of the quark propagator~\cite{Faccioli:2003qz,
  Bowman:2004xi} are in a remarkable agreement with that of
Ref.~\cite{Diakonov:1983hh}. A recent lattice simulation 
with the interacting instanton liquid model obtains $\bar
\rho\simeq 0.32\,\mathrm{fm}$ and $\bar R\simeq 0.76 
\,\mathrm{fm}$ with the finite current quark mass $m$ taken into
account~\cite{Cristoforetti:2006ar}.

The smallness of the packing parameter $\pi \bar{\rho}^4/ \bar{R}^4
\approx 0.1$ makes it possible to average the determinant 
over collective coordinates of instantons with fermionic
quasi-particles, {\it i.e.} constituent quarks $\psi$ introduced. The
averaged determinant turns out to be the light-quark partition
function $Z[V,m]$ which is a functional of the electromagnetic field
$V$ and can be represented by a functional integral over the
constituent quark fields with the gauged effective chiral action 
$S[\psi^\dagger,\psi,V]$.  However, it is not trivial to make
the action gauge-invariant due to the nonlocality of the
quark-quark interactions generated by instantons. In the previous 
paper~\cite{Musakhanov:2002xa}, it was demonstrated how to gauge
the nonlocal effective chiral action in the presence of the
external electromagnetic field and was shown that the low-energy
theorem of the axial anomaly relevant to the process $G\tilde
G\rightarrow \gamma\gamma$ is satisfied.  Moreover, the gauged
effective chiral action was shown to be very successful in describing
various observable of mesons and vacuum
properties~\cite{Nam:2006ng, Kim:2004hd, Nam:2006sx, Ryu:2006bf,
Nam:2007fx, Goeke:2007bj, Goeke:2007nc}.    
   
In the present work, we want to show that the pion EM form factor is
indeed in good agreement with experimental data.  In addition, the 
results for the kaon EM form factor will be given.  The EM charge
radii of the mesons are shown to be: $\langle
r^2\rangle_{\pi^+}=0.455\,\mathrm{fm}^2$, $\langle
r^2\rangle_{K^+}=0.486\,\mathrm{fm}^2$ and $\langle
r^2\rangle_{K^0}=-0.060\,\mathrm{fm}^2$ without any 
adjustable free parameter, which are all in good agreement with 
experimental data. The low-energy constant $L_9$ in the large $N_c$
limit is found to be $8.42\times10^{-3}$.

We organize the present work as follows: In Section II, we briefly 
explain the general formalism relevant for studying the EM form 
factors.  In Section III, we introduce the nonlocal chiral quark model 
from the instanton vacuum.  In Section IV, the numerical results are  
discussed and compared with those of other works.  The final 
Section is devoted to summarize the present work and to draw  
conclusions.  

\section{Electromagnetic form factors for pion and kaon}
The electromagnetic (EM) form factor of a pseudoscalar meson
$F_{\pi,K}$ can be defined by the following matrix element: 
\begin{equation}
 \label{eq:emff}
\langle \mathcal{M} (P_f)|j_{\mu}^{\rm EM}(0) |\mathcal{M} (P_i)\rangle =
(P_i+P_f)_{\mu}F_{\mathcal{M}} (q^2),
\end{equation}
where $|\mathcal{M}\rangle$ stands for the pseudoscalar meson state,
for example, $|\pi^+(u\bar{d})\rangle$, 
$|K^0(d\bar{s})\rangle$ and $|K^+(u\bar{s})\rangle$. The $P_i$ and
$P_f$ represent the initial and final 
on-shell momenta for the meson, respectively. Then the masses of the
mesons can be obtained from $P^2_i=P^2_f=m^2_{\pi,K}$, and we use
$140$ MeV and $495$ MeV for the pion and kaon masses,
respectively. The momentum transfer is 
written as $q=P_f-P_i$, of which the square is given by $-q^2=Q^2>0$ in
the space-like region.  Here, the EM current in flavor SU(3) is 
defined as follows:   
\begin{equation}
\label{eq:emcurrent}
j^{\rm EM}_{\mu}(x) = \sum_{\mathrm{u,d,s}} \, e_f \psi_f^{\dagger}(x)
\gamma_{\mu} \psi_f(x)
=\frac{2}{3}u^{\dagger}(x)\gamma_{\mu}u(x)-\frac{1}{3} 
d^{\dagger}(x)\gamma_{\mu}d(x)-\frac{1}{3}s^{\dagger}(x)\gamma_{\mu}s(x),
\end{equation}
where $e_f$ indicates the electric charge of the quark with flavor
$f$.  Note that all calculations are performed in Euclidean space,
since we are interested in the form factor in the space-like region.
The EM form factors satisfy the normalization conditions at $Q^2=0$ by 
the gauge invariance:  
\begin{equation}
\label{eq:norm}
F_{\pi^+}(0)=F_{K^+}(0)=1,\,\,\,\,F_{K^0}(0)=0.
\end{equation}

The EM (vector) mean square charge radii for the mesons can be
evaluated by the derivatives of the form factors at $Q^2=0$:    
\begin{equation}
  \label{eq:radi}
\langle r^2\rangle_{\mathcal{M}} = -6\left[\frac{\partial\,F_{\mathcal{M}} (Q^2)}
{\partial\,Q^2}\right]_{Q^2=0}.
\label{radius}
\end{equation}
Moreover, the pion charge radius relates to the Gasser-Leutwyler 
low-energy constant (LEC) $L_9$~\cite{Gasser:1983yg,Bijnens:2002hp} in
the large $N_c$ limit:  
\begin{equation}
\label{eq:l9}
\langle r^2\rangle_{\pi^+}=\frac{12L_9}{F^2_{\pi}}.
\end{equation}
\section{Gauged effective chiral action in the presence of
  the electromagnetic field}
The light quark partition function $Z[V,m]$ with the current quark
mass $m$ and in the presence of the external vector field $V_\mu$ is
defined as 
\begin{equation}
\label{eq:lqpar}
Z[V,m] = \int DA_\mu e^{-G^4/4} \mathrm{Det}\left[i\rlap{/}{\partial} +
  \rlap{/}{A} +  \rlap{/}{V}+im \right],  
\end{equation}
where $A_\mu$ denotes the gluon field and $G_{\mu\nu}$ is the gluon
field strength tensor.  The $m$ stands for the current quark mass.  
In the instanton liquid model, it is assumed that the functional
integral of Eq.~(\ref{eq:lqpar}) can be solved by expanding it around
the classical vacuum.  It was already shown in 
Refs.~\cite{Diakonov:1983hh,Diakonov:1985eg} how to solve the integral
in Eq.~(\ref{eq:lqpar}) in the chiral limit and in the absence of the
external fields.  In the present work, we briefly review a method 
how to derive an low-energy effective partition function solve this
integral in the presence of the external vector field, closely
following Refs.~\cite{Musakhanov:2002xa,Kim:2004hd}.    

The quark determinant in Eq.~(\ref{eq:lqpar}) can be decomposed into 
the low- and high-frequency parts: $\mathrm{Det} = \mathrm{Det_{low}}
\cdot \mathrm{Det_{high}}$.  While the high-frequency part is related
to the perturbative renormalization, the low-frequency one, which we
consider here, is relevant to the low-energy regime.  Thus, we
consider here the low-frequency part.  We first consider the zero-mode
approximation for the quark propagator interacting with the $i$-th
instanton: 
\begin{equation} 
S_{i} = \frac{1}{i\rlap{/}{\partial} + \rlap{/}{A}_i + im} =
\frac{1}{\rlap{/}{p}} +
\frac{|\Phi_{i,0}\rangle\langle\Phi_{i,0}|}{im},
\end{equation} 
where $|\Phi_{i,0}\rangle$ represents the quark zero-mode solution,
satisfying 
\begin{equation}
\label{eq:QZMS}
\left(i\rlap{/}{\partial} + \rlap{/}{A}_i \right)|\Phi_{i,0}\rangle=0.
\end{equation}
Though this zero-mode approximation works well for the very small
mass of the quark, we have to extend it in order to consider the
finite current quark mass as follows:
\begin{equation}  
\label{Si}
S_i=S_{0} + S_{0}\rlap{/}{p}\frac{|\Phi_{0i}\rangle\langle
\Phi_{0i}|}{c_i}  \rlap{/}{p} S_{0} ,
\end{equation}
where the $S_0$ is the free quark propagator:
\begin{equation}
\label{feeprop}
S_0=\frac{1}{\rlap{/}{p} +im}  
\end{equation}
and $c_i$ is a matrix element defined as 
\begin{equation}
c_i=-\langle\Phi_{0i}|\rlap{/}{p} S_{0} \rlap{/}{p} |\Phi_{0i}\rangle
= i m\langle\Phi_{0i}|S_{0}\rlap{/}{p} |\Phi_{0i}\rangle = im
\langle\Phi_{0i}|\rlap{/}{p} S_{0}|\Phi_{0i}\rangle.
\end{equation} 
The approximation given in Eq.~(\ref{Si}) allows us to project
$S_i$ to the correct zero-modes with the finite $m$:
\begin{equation}
S_i|\Phi_{0i}\rangle = \frac{1}{im}|\Phi_{0i}\rangle,\,\,\,
\langle\Phi_{0i}|S_i =\langle\Phi_{0i}|\frac{1}{im}.
\end{equation}

Now, we introduce the external vector field, so that the quark
propagator is modified as 
\begin{equation}
\tilde S = \frac{1}{\rlap{/}{p} + \rlap{/}{A} +\rlap{/}{V} +
im},\;\; \tilde S_i= \frac{1}{\rlap{/}{p} + \rlap{/}{A_i}
+\rlap{/}{V}  +im}.
\end{equation}
Here, we assume that the total instanton field $A$ may be
approximated as a sum of the single instanton fields,
$A=\sum_{i=1}^N A_i$, which is justified in the dilute instanton liquid
($\bar{\rho}\simeq\frac{1}{3}$ fm and $\bar{R}\simeq 1$ fm).  If we switch off the
instanton fields, the we can express the quark propagator in the
presence of the external fields as follows:  
\begin{equation}
\tilde S_0=\frac{1}{\rlap{/}{p} + \rlap{/}{V} + im}. 
\end{equation}
The total propagator $\tilde{S}$ can be now expanded with respect to a
single instanton $A_i$:  
\begin{equation}
\label{S-tot} 
\tilde S=\tilde S_0+\sum_i (\tilde S_i-\tilde S_0)+\sum_{i\not=j} (\tilde
S_i-\tilde S_0)\tilde S^{-1}_0(\tilde S_j-\tilde S_0)+\cdots .  
\end{equation}
Expanding $\tilde{S}$ with respect to the external field $V$, we are
able to express $\tilde S_i$ in terms of $S_i$.  Since we use the
zero-mode approximation, the gauge invariance for the external vector
field is broken.  Thus, we have to restore it by introducing the
following auxiliary field $V^{\prime }$ and gauge connection $L_i$:
\begin{equation}
\rlap{/}{V}_i^\prime=\bar L_i(\rlap{/}{p}+\rlap{/}{V})L_i-\rlap{/}{p}, 
\end{equation}
where the gauge connection $L_i$ is defined as a path-ordered exponent  
\begin{equation}
\label{transporter}
L_i(x,z_i)=\mathrm{P} \exp\left[i\int_{z_i}^x ds_\mu V_\mu(s)\right],
\,\,\,\bar L_i(x,z_i) =\gamma_0 L_i^\dagger(x,z_i)\gamma_0  
\end{equation}
with an instanton coordinate $z_i$.  The field $V^{\prime }_i(x,z_i)$
under flavor rotation $\psi(x)\rightarrow U(x)\psi(x)$ is transformed
as $V^{'}_i(x,z_i)\rightarrow U(z_i)V^{\prime }_i(x,z_i)U^{-1}(z_i)$. The
propagators $\tilde S_i$ and $\tilde S_0$ then have the following
form:
\begin{equation}
\label{eq:a}
\tilde{S}_i = L_iS^{\prime}_{i}\bar L_i,\,\,\, S^{\prime}_{i}
=\frac{1}{\rlap{/}{p} +\rlap{/}{A_i}+\rlap{/}{V_i^{'}}+im},\,\,\,
\tilde{S}_0 = L_iS^{\prime}_{0i}\bar L_i,\;\; S^{\prime }_{0i}
=\frac{1}{\rlap{/}{p} + \rlap{/}{V_i^{\prime}}+im},
\end{equation}
Expanding $S^{\prime }_{i}$ with respect to $\rlap{/}{V}_{i}^{\prime }$ and
resumming it, we obtain the following expression:
\begin{equation}
\label{eq:b}
S^{\prime }_{i} = S_i\left[1+\sum_n (-\hat V_{i}^{\prime }S_i)^n\right]=S^{\prime
}_{0i} + S^{\prime }_{0i}\rlap{/}{p}\frac{|\Phi_{0i}\rangle\langle \Phi_{0i}|
}{c_i - b_i} \rlap{/}{p} S^{\prime }_{0i} ,  \label{Si'}
\end{equation}
where 
\begin{equation}
\label{eq:bi}
b_i =\langle\Phi_{0i}|\rlap{/}{p} (S^{\prime}_{0i}-S_0) \rlap{/}{p}
|\Phi_{0i}\rangle,\,\,\,
c_i - b_i =-\langle\Phi_{0i}| \rlap{/}{p} S^{\prime}_{0i} \rlap{/}{p}
|\Phi_{0i}\rangle. 
\end{equation}
Rearranging Eq.~(\ref{S-tot}) for the total propagator, we get 
\begin{eqnarray}
\tilde S&=&\tilde S_{0} + \tilde S_{0}\sum_{i,j} \bar L_i^{-1} \rlap{/}{p}
|\Phi_{i0}\rangle\left[\frac{1}{-\mathcal{D}}+\frac{1}{-\mathcal{D}}
\mathcal{C} \frac{1}{-\mathcal{D}} +  \cdots\right]_{ij} \langle\Phi_{0j}|
\rlap{/}{p} L_j^{-1}\tilde S_{0}\cr &=&\tilde S_{0} + \tilde
S_{0}\sum_{i,j}\bar L_i^{-1}\rlap{/}{p} |\Phi_{i0}\rangle
\left[\frac{1}{-\mathcal{V}-\mathcal{T}}\right]_{ij}
\langle\Phi_{0j}|\rlap{/}{p} L_j^{-1}\tilde S_{0} ,  
\label{propagator}
\end{eqnarray}
where
\begin{eqnarray}
\label{eq:c}
\mathcal{V}_{ij} &=& \langle\Phi_{0i}|\rlap{/}{p} (L_i^{-1}\tilde S_{0}\bar
L^{-1}_j) \rlap{/}{p}|\Phi_{0j}\rangle -\langle\Phi_{0i}|\rlap{/}{p} S_{0}
L_i^{-1}L_j \rlap{/}{p} |\Phi_{0j}\rangle,  \cr
\mathcal{T}_{ij}&=&(1-\delta_{ij})\langle\Phi_{0i}|\rlap{/}{p} S_0
L_i^{-1}L_j\rlap{/}{p} |\Phi_{0j}\rangle,  \cr
\mathcal{D}_{ij}&=&\delta_{ij} \mathcal{V}_{ij} \equiv (b_i-c_i)
\delta_{ij}, \cr \mathcal{C}_{ij}&=&(1-\delta_{ij})\mathcal{V}_{ij}.
\end{eqnarray}
We introduce now the modified zero-mode solution as follows:
\begin{equation}
|\phi_0\rangle=\frac{1}{\rlap{/}{p}}L \rlap{/}{p} |\Phi_0\rangle,
\end{equation}
which has the same chiral properties as the zero-mode solution $
|\Phi_0\rangle$.  Then we get
\begin{equation}
\tilde S -\tilde S_{0} = -\tilde S_{0}\sum_{i,j}\rlap{/}{p} | \phi_{0i}
\rangle \langle\phi_{0i}|\frac{1}{\mathcal{V}+\mathcal{T}}|
\phi_{0j} \rangle \langle\phi_{0j}|\rlap{/}{p} \tilde S_{0}
\label{propagator1}
\end{equation}
with
\begin{equation}
\mathcal{V}+\mathcal{T}=\rlap{/}{p} \tilde S_{0}\rlap{/}{p}.
\end{equation}
The final expresion for Eq.~(\ref{propagator1}) is obtained as
\begin{equation}
\mathrm{Tr} (\tilde S -\tilde S_0 )= -\sum_{i,j}\langle\phi_{0,j}|\rlap{/}{p}
\,{\tilde S_{0}}^2\, \rlap{/}{p} |\phi_{0,i}\rangle\langle\phi_{0,i}|
\frac{1}{ \rlap{/}{p}\tilde S_{0}\rlap{/}{p}}|\phi_{0,j}\rangle.
\end{equation}

In order to derive the low-frequency part of the quark
determinant, we now introduce a matrix operator $\tilde{B}(m)$ defined
as follows:
\begin{equation}
\tilde B(m)_{ij} = \langle\phi_{0,i}|\rlap{/}{p} \tilde S_0 \rlap{/}{p}
|\phi_{0,j}\rangle 
= \langle\Phi_{0i}|\rlap{/}{p}\left[L_{i}^{-1} \,\tilde S_{0}\bar
  L_{j}^{-1}\right]\rlap{/}{p} |\Phi_{0j}\rangle,
\end{equation}
where $i,j$ are indices for the different instantons. Then, we can
show that
\begin{eqnarray}
\ln{\left(\mathrm{Det}_{\mathrm{low}}\right)}& = & 
\mathrm{Tr} \ln{\left[
\frac{i\rlap{/}{\partial} + \rlap{/}{A} + \rlap{/}{V}+ im}
{i\rlap{/}{\partial}+\rlap{/}{V}+ im} \right]}= i \mathrm{Tr} 
\int^m dm^{\prime }\left[\tilde S(m^{\prime })- \tilde
S_{0}(m^{\prime })\right]\cr &=& \sum_{i,j} \int^m dm' 
\frac{d \tilde B(m^{\prime})_{ij}}{dm'}
\left[\tilde B(m')\right]_{ji}^{-1} =\mathrm{Tr} \ln {\tilde
B(m)},  \nonumber
\end{eqnarray}
where ${\rm Tr}$ denotes the trace over the subspace of the quark zero
modes. Thus, we have 
\begin{equation}
{\mathrm{Det}}_{\mathrm{low}} [V,m] \cong \mathrm{Det} \tilde
B(m),  \label{tildeB}
\end{equation}
where $\tilde B$ is identified as an extension of the Lee-Bardeen
matrix $B$~\cite{Lee:1979sm} in the presence of the external vector
field $V$.  Taking $m$ to be small and switching off the external
fields, we can show that $\tilde B$ turns out to be the same as $B$ to
order $\mathcal{O}(m)$. 

Averaging $\mathrm{Det}_{\mathrm{low}}$ over the instanton
collective coordinates $z$, which provides the effective chiral partition
function $Z[V,m]$, we can obtain the fermionized representation of 
the partition function in Eq.~(\ref{tildeB}): 
\begin{equation}
\label{eq:PF1}
Z _{\rm eff}[V,m] 
= \int D\psi D\psi^{\dagger} \exp\left[\int d^4 x \psi^{\dagger}
(i\rlap{/}{D}+ im)\psi\right] \prod^{N_{\pm}}
W_{\pm}[\psi^{\dagger} ,\psi ],  
\end{equation}
where $iD_\mu$ is the covariant derivative defined as   
\begin{equation}
  \label{eq:covd1}
 i D_\mu = i\partial_\mu + V_\mu.
\end{equation}
$m$ represents the current quark mass:
$m=\mathrm{diag}(m_{\mathrm{u}}, m_{\mathrm{d}},m_{\mathrm{s}})$, and
the quark-instanton interaction, $W_{\pm}$ is given by 
\begin{eqnarray}
\label{eq:tildeV}
&&W_{\pm}[\psi^{\dagger} ,\psi]
\nonumber\\
&&=\int d^4z_{\pm}\prod_{N_f}\int
d^4x\,d^4y \left[\psi^{\dagger} (x)\bar L^{-1}(x,z_\pm)
\rlap{/}{p} \Phi_{\pm,0} (x; z_{\pm})\right]
\left[\Phi_{\pm , 0}^\dagger (y; z_{\pm} ) (\rlap{/}{p}\,
L^{-1}(y,z_\pm) \psi (y)\right]. 
\end{eqnarray}
The fermion fields $\psi^{\dagger}$ and $\psi$ are interpreted as the
dynamical quark fields or constituent quark fields induced by the
zero-modes of the instantons.

Note that the partition function of Eq.~(\ref{eq:PF1}) is
invariant under local flavor rotations due to the gauge connection
$L$ in the interaction term $W_{\pm}[\psi^{\dagger} ,\psi ]$.  While
we preserve the gauge invariance of the effective chiral action by
introducing the gauge connection, we have to pay a price: The
effective action depends on the path in the gauge connection $L$. We
will choose the straight-line path as a trial for brevity, though
there is in general no physical reason why other choices should be
excluded.  

The low-energy partition function given in Eq.~(\ref{eq:PF1}) can be
bosonized as shown in Refs.~\cite{Diakonov:1983hh,Diakonov:2002fq}.
While the bosonization procedure is exact in the case of $N_f=2$, it
is nontrivial for $N_f\ge3$.  However, as pointed out in
Refs.~\cite{Diakonov:1983hh,Diakonov:1997sj,Diakonov:2002fq}, one
still can do it taking the limit of large $N_c$. 
Assuming that the external vector field is weak and taking the large
$N_c$ limit, we can express the quark-instanton interaction in
Eq.~(\ref{eq:PF1}) in the following simpler form: 
\begin{eqnarray}
\label{eq:WJ}
W_{\pm}[\psi^{\dagger},\psi]&=&(-i)^{N_f}
\left[\frac{(2\pi\bar{\rho})^2}{N_c}\right]^{N_f}
\int \frac{d^4x}{V} \,{\rm det}_f[iJ_{\pm}(x)],
\nonumber\\
J_{\pm}(x)&=&\int\frac{d^4k\,d^4p}{(2\pi)^8}
e^{i(k-p)\cdot x}[2\pi\bar{\rho}F(K)]
[2\pi\bar{\rho}F(P)]\left[\psi^{\dagger}(k)
  \frac{1\pm\gamma_5}{2}\psi(p)\right], \,\,\,\,N_c\to\infty, 
\end{eqnarray}
where we have dropped the color and flavor indices. Note that
$J_{\pm}$ is identified by its chirality. The form factor $F(k)$ is
the Fourier transform of the quark zero-mode solution: 
\begin{equation}
\label{eq:ZZZZ}
F(k)=2t\left[I_0(t)K_1(t)-I_1(t)K_0(t)-\frac{1}{t}I_1(t)K_1(t)\right],
\,\,\,\, t=\frac{k\bar{\rho}}{2}.
\end{equation}
The $V$ represents the volume factor of Euclidean space.
Note that the quark-instanton interaction given in Eq.~(\ref{eq:WJ}) is 
simply the same form as the original one with the gauged form factor
$F(K)$.  Here, the $K$ denotes the covariant momentum $k_\mu-  
V_\mu$.  Thus, the low-energy effective partition function can be
written as 
\begin{equation}
  \label{eq:pf12}
Z_{\mathrm{eff}}[V,m] = \int D\psi D\psi^\dagger \exp\left(\int d^4 x \sum_f
  \psi_f^\dagger (i\rlap{/}{D} + im)\psi_f\right)
W_{+}^{N_+} W_-^{N_-}. 
\end{equation}

The detailed procedure of the bosonization is found in
Refs.~\cite{Diakonov:1997sj}.  Thus, having bosonized
Eq.~(\ref{eq:pf12}), we arrive at the following expression for the
low-energy effective partition function:
\begin{equation}
\label{eq:ECA}
Z_{\rm eff}[\mathcal{M}^{\alpha},V,m] =\int D\mathcal{M}^{\alpha}
\exp\left(-\mathcal{S}_{\rm eff}  [\mathcal{M}^{\alpha},V,m]
\right),   
\label{Z}
\end{equation}
where $\mathcal{S}_{\rm eff}$ stands for the effective chiral
action 
\begin{equation}
\label{eq:ECA1}
\mathcal{S}_{\rm eff}[\mathcal{M}^{\alpha},V,m]  = -{\rm Sp}_{c,f,\gamma} \ln
\left[i\rlap{/}{D}+im + i\sqrt{M (iD)}  U^{\gamma_5}\sqrt{M
  (iD)}\right]. 
\end{equation}
Here, the ${\rm Sp}_{c,f,\gamma} $ stands for the functional trace,
i.e.  $\int d^4x\,{\rm Tr}_{c,f,\gamma}\langle x|\cdots|x\rangle$, in which
the subscripts $c$, $f$ and $\gamma$ represent the traces over color,
flavor and Dirac spin spaces, respectively. The $U^{\gamma _{5}}$
reads:    
\begin{equation}
\label{eq:d}
U^{\gamma _{5}}=U(x)\frac{1+\gamma _{5}}{2}+U^{\dagger }(x)
\frac{1-\gamma_{5}}{2}
=1+\frac{i}{F_{\pi}}\gamma _{5}(\mathcal{M}\cdot
\lambda)-\frac{1}{2F_{\pi}^{2}}(\mathcal{M}\cdot\lambda)^2\cdots.
\end{equation}
The pseudoscalar meson fields $\pi^{\alpha}$ are defined explicitly as
\begin{eqnarray}
\mathcal{M}\cdot \lambda&=&\frac{1}{\sqrt{2}}\left(
\begin{array}{ccc}
\frac{1}{\sqrt{2}}\pi^0+\frac{1}{\sqrt{6}}\eta&\pi^+&K^+\\
\pi^-&-\frac{1}{\sqrt{2}}\pi^0+\frac{1}{\sqrt{6}}\eta&K^0\\
K^-&\bar{K}^0&-\frac{2}{\sqrt{6}}\eta\\
\end{array}\right),
\end{eqnarray} 
where $\lambda^{\alpha}$ stands for the Gell-Mann SU(3) flavor matrix,
satisfying
$[\lambda^{\alpha},\lambda^{\beta}]=2\delta^{\alpha\beta}$. 

The dynamical quark mass in the square brakets in the {\it r.h.s.} of 
Eq.~(\ref{eq:ECA}) can be written explicitly as follows:    
\begin{equation}
\label{eq:ff0}
M(iD)=M_0F^2(iD)f(m)=M_0F^2(iD)\left[\sqrt{1+\frac{m^2}{d^2}}
-\frac{m}{d} \right],   
\end{equation}
where $F(iD)$ denotes the form factor arising from the Fourier
transform of the quark zero-mode solution and consists of the modified
Bessel functions of the second kind as shown in
Eq.~(\ref{eq:ZZZZ}). However, in the present calculation, 
we take a simple dipole-type parameterization of
$F(iD)$ for numerical simplicity:
\begin{equation}
\label{eq:ff1}
F(iD)=\frac{2\Lambda^2}{2\Lambda^2-D^2}, 
\end{equation}
where $\Lambda$ is the cut-off mass being chosen to be
$1/{\bar\rho}\approx 600$ MeV as usual. The curent-quark mass ($m$)
correction term, $f(m)$ in Eq.~(\ref{eq:ff0}) is obtained as in
Refs.~\cite{Pobylitsa:1989uq,Musakhanov:1998wp}.  The parameter $d$ 
can be calculated uniquely as follows~\cite{Musakhanov:2001pc}:    
\begin{equation}
\label{eq:dd}
d=\sqrt{\frac{0.08385}{2N_c}}\frac{8\pi\bar{\rho}}{R^2}\approx0.198\,{\rm 
  GeV}.   
\end{equation}
The value of the constituent quark mass at zero momentum $M_0$ can be
self-consistently determined by the saddle-point equation:
\begin{equation}
\label{eq:saddle}
\frac{N}{V}
=4N_c\int\frac{d^4k}{(2\pi)^4}\frac{M^2(k)}{k^2+M^2(k)},
\end{equation}
which gives $M_0\approx350$ MeV.  We select the up and down quark masses
to be $5$ MeV and the strange quark mass to be $150$ MeV.     
\section{Electromagnetic formfactors for pion and kaon from the
  instanton vacuum}
\begin{figure}[t]
\includegraphics[width=12cm]{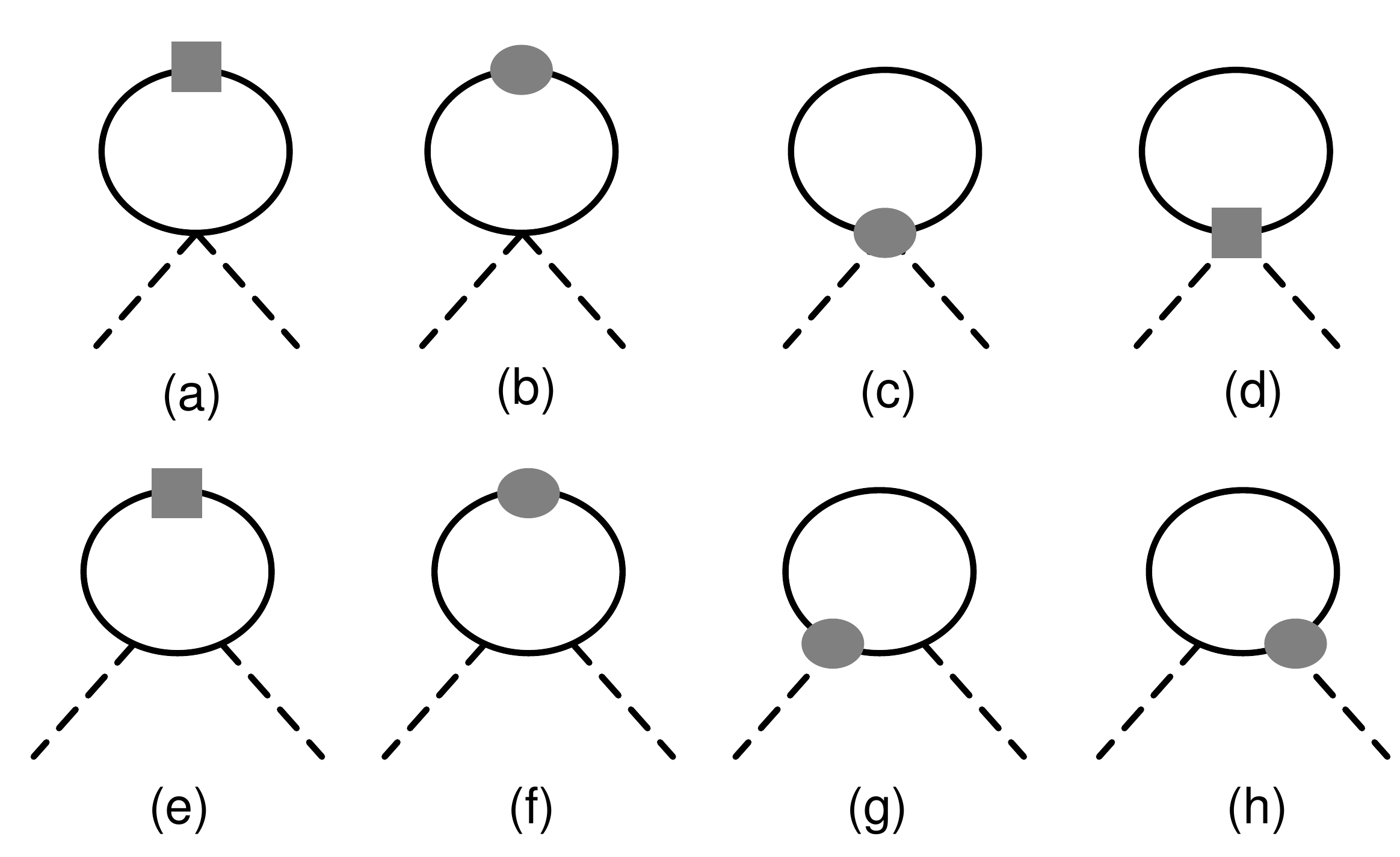}
\caption{Relevant Feynman diagrams for the pion (kaon)
  EM form factor. The solid and dashed lines denote the quark and
  pseudoscalar meson, respectively.  The solid box and circle stand
  for the local- and  nonlocal-interaction vertices, respectively, to
  which the external 
photon field is coupled.}        
\label{fig:1}
\end{figure}
We are now in a position to calculate the matrix element for the EM 
form factor in Eq.~(\ref{eq:emff}) using the gauged effective chiral
action, defined in Eq.~(\ref{eq:ECA1}). It is worth mentioning that
all relevant diagrams for the EM form factor are obtained from the
gauged effective chiral action straightforwardly.  Moreover, the
Ward-Takahashi identity is fully satisfied as well, that is, the pion
EM form factor at $Q^2=0$ becomes exactly $F_\pi(0)=1$.  The meson
matrix element in Eq.~(\ref{eq:emff}) can be evaluated via the
functional derivative of the effective chiral 
action:   
\begin{eqnarray}
\label{eq:FD}
&&\left(\frac{\delta^3 \mathcal{S}_\mathrm{eff}[\pi,V,m]}
{\delta{\pi}^{\alpha}(y)\,
\delta{\pi}^{\beta}(x)\,\delta V_{\mu}(0)}\right)_{V=0}
=e_qN_c{\rm Tr}_{f,\gamma}
\Bigg[\underbrace{\frac{1}{\mathcal{D}}X^{{\alpha}{\beta}}\frac{1}{\mathcal{D}}
\left(\gamma_{\mu} +W_{\mu}-Z_{\mu}\right)}_{\rm a,b} 
-\underbrace{\frac{1}{\mathcal{D}}\left(W^{{\alpha}{\beta}}_{\mu}-
    Z^{{\alpha}{\beta}}_{\mu}\right)}_{\rm c} 
\nonumber\\
&&\hspace{2cm}\underbrace{
\frac{1}{\mathcal{D}}X^{\alpha}\frac{1}{\mathcal{D}}X^{\beta}
\frac{1}{\mathcal{D}}
\left(\gamma_{\mu} +W_{\mu}-Z_{\mu}\right) 
\frac{1}{\mathcal{D}}X^{\beta}\frac{1}{\mathcal{D}}X^{\alpha}
\frac{1}{\mathcal{D}}
\left(\gamma_{\mu} +W_{\mu}-Z_{\mu}\right)}_{\rm e,f}
\nonumber\\
&&\hspace{3cm}
+\underbrace{\frac{1}{\mathcal{D}}X^{\alpha}\frac{1}{\mathcal{D}}
\left(W^{\beta}_{\mu}-Z^{\beta}_{\mu}\right)}_{\rm g}
+\underbrace{\frac{1}{\mathcal{D}}X^{\beta}\frac{1}{\mathcal{D}}
\left(W^{\alpha}_{\mu}-Z^{\beta}_{\mu}\right)}_{\rm h}
\Bigg],
\end{eqnarray}
where we use the following abbreviations:
\begin{equation}
\label{eq:notations}
\mathcal{D}=i\rlap{/}{\partial}+im+iM(i\partial),
\,\,\,\,
\frac{\delta X}{\delta\pi^{\alpha}}=X^{\alpha},
\,\,\,\,
\frac{\delta^2 X}{\delta\pi^{\alpha}\delta\pi^{\beta}}=X^{\alpha\beta}
\end{equation}
with
\begin{equation}
X=i\sqrt{M(i\partial)}U^{\gamma_5}\sqrt{M(i\partial)},
\end{equation}
and 
\begin{equation}
W_{\mu}=i\left(\frac{\partial}{\partial p_\mu}\sqrt{M(i\partial)}\right)
U^{\gamma_5}\sqrt{M(i\partial)}, \,\,\,\,
Z_{\mu}=i\sqrt{M(i\partial)}U^{\gamma_5}\left(\frac{\partial}{\partial
  p_\mu}\sqrt{M(i\partial)}\right). 
\end{equation}
In Eq.~(\ref{eq:FD}) we group each term corresponding to
Fig.~\ref{fig:1}, in which we show all relevant Feynman
diagrams containing local (solid box) and nonlocal (solid circle)
interacting vertices to the photon field.  The solid and dashed lines
represent the quark and meson fields, respectively.  The terms with
$V$ and $W$ represents those with the nonlocal interacting vertices.
Note that, within the present model, diagram (d), which contains
one-quark loop and a local interacting vertex, is not generated as
shown in Eq.~(\ref{eq:FD}).    

Using Eqs.~(\ref{eq:FD}), (\ref{eq:notations}) and carrying out a
straightforward manipulation of the functional trace, we have seven 
contributions to the meson EM form factor as depicted
schematically in Fig.~\ref{fig:1} excluding 
diagram (d).  Diagrams (a), (b) and (c) stand for the isosinglet
contributions, whereas all other diagrams contain both the isosinglet
and isotriplet ones.  Note that diagram (e) arises from the
three-point correlation function in the (local) impulse
approximation.  Diagram (c), however, becomes zero, since they
contain a single quark propagator.  We list the properties of all
these diagrams in Table~\ref{table00} as summary. 
\begin{table}[b]
\begin{tabular}{c||c|c|c|c|c|c|c|c}
Diagram&(a)&(b)&(c)&(d)&(e)&(f)&(g)&(h)\\
\hline
Interaction&L&NL&NL&L&L&NL&NL&NL\\
\hline
Isospin&S&S&S&S&S,T&S,T&S,T&S,T\\
\hline
Contribution&$\circ$&$\circ$&$\times$&$\times$&$\circ$&
$\circ$&$\circ$&$\circ$\\
\end{tabular}
\caption{Properties of the relevant diagrams.  We use the following 
abbreviations: L (local), NL (nonlocal), S (singlet) and T (triplet).}  
\label{table00}
\end{table}
Having performed all remained calculations, we arrive at the
final expression for the pion and kaon EM form factors:  
\begin{equation}
  \label{eq:EMFF}
F_{\pi,K}(Q^2)=\frac{1}{F^2_{\pi,K}}\sum_{\rm flavor}
\frac{8e_qN_c}{(2p_i\cdot q+m^2_{\pi,K})}
\int^{\infty}_{-\infty}\frac{d^4k}{(2\pi)^4}\sum_{i={\rm a}}^{\rm h}
\mathcal{F}_i(k,q),
\end{equation}
where
\begin{eqnarray}
\label{eq:EMFF16}
\mathcal{F}^{\rm L}_{\rm a}&=&
\frac{\sqrt{M_bM_c}(\bar{M}_ck_{bd}+\bar{M}_bk_{cd})}
{2(k^2_b+\bar{M}^2_b)(k^2_c+\bar{M}^2_c)},
\nonumber\\
\mathcal{F}^{\rm NL}_{\rm b}&=&
\frac{\sqrt{M_bM_c}(\sqrt{M_c}\hat{M}_{bd}-\sqrt{M_b}\hat{M}_{cd})
(k_{bc}-\bar{M}_b\bar{M}_c)}{(k^2_b+\bar{M}^2_b)(k^2_c+\bar{M}^2_c)},
\,\,\,\,\mathcal{F}^{\rm NL}_{\rm c}=0,
\,\,\,\,\mathcal{F}^{\rm NL}_{\rm d}=0,
\nonumber\\
\mathcal{F}^{\rm L}_{\rm e}&=&
\frac{M_a\sqrt{M_bM_c}(k_{ab}k_{cd}+k_{ac}k_{bd}
-k_{bc}k_{ad}+\bar{M}_a\bar{M}_ck_{bd}+\bar{M}_a\bar{M}_bk_{cd} 
-\bar{M}_b\bar{M}_ck_{ad})}{(k^2_a+\bar{M}^2_a)(k^2_b+\bar{M}^2_b)
(k^2_c+\bar{M}^2_c)},
\nonumber\\
\mathcal{F}^{\rm NL}_{\rm f}&=&
\frac{M_a\sqrt{M_bM_c}(\sqrt{M_b}
\hat{M}_{cd}-\sqrt{M_c}\hat{M}_{bd})(\bar{M}_ck_{ab} +
\bar{M}_bk_{ac}-\bar{M}_ak_{bc}+\bar{M}_a\bar{M}_b\bar{M}_c)}  
{(k^2_a+\bar{M}^2_a)(k^2_b+\bar{M}^2_b)(k^2_c+\bar{M}^2_c)},
\nonumber\\
\mathcal{F}^{\rm NL}_{\rm g}&=&
\frac{\sqrt{M_aM_c}
\left[\sqrt{M_b}\hat{M}_{ad}-\sqrt{M_a}\hat{M}_{bd}\right]
(k_{ac}+\bar{M}_a\bar{M}_c)}
{2(k^2_a+\bar{M}^2_a)(k^2_c+\bar{M}^2_c)},
\nonumber\\
\mathcal{F}^{\rm NL}_{\rm h}&=&
\frac{\sqrt{M_aM_b}
\left[\sqrt{M_c}\hat{M}_{ad}-\sqrt{M_a}\hat{M}_{cd}\right]
(k_{ab}+\bar{M}_a\bar{M}_b)}
{2(k^2_a+\bar{M}^2_a)(k^2_b+\bar{M}^2_b)},
\end{eqnarray}
where the subscripts $a\sim h$ correspond to the diagrams shown in
Fig.~\ref{fig:1}. Note that all integrals are carried out 
in Euclidean space.  The notations are defined as $k_a=k-P_i/2-q/2$,
$k_b=k+P_i/2-q/2$, $k_c=k+P_i/2+q/2$ and $k_d=P_i$. The
$\bar{M}_{\alpha}$ is the sum of the current and momentum-dependent
quark masses, $m+M(k_{\alpha})$. We use also the following
abbreviations:   
$k_{\alpha\beta}=k_{\alpha}\cdot k_{\beta}$, and
$\hat{M}_{\alpha\beta}$ is defined as the  derivative of the 
dynamical-quark mass with respect to the momenta: 
\begin{equation}
\hat{M}_{\alpha\beta}=\frac{\partial\sqrt{M_{\alpha}}}{\partial
  k^{\mu}_{\alpha}}k_{\beta\mu}=-\frac{4\sqrt{M}\Lambda^2}
  {(2\Lambda^2+k^2_{\alpha})^2} (k_{\alpha}\cdot k_{\beta}).
\end{equation}
\section{Numerical results and discussion}
In this Section, we discuss the numerical results for the pion and 
kaon EM form factors.  Since our framework is fully relativistic, we 
choose the Breit-momentum frame for convenience as done in
Refs.~\cite{Roberts:1994hh,Nam:2007fx} (see Appendix).  In 
the upper-left panel of Fig.~\ref{fig:1}, we draw each contribution
from the diagrams depicted in Fig.~\ref{fig:1}. Note that the
diagrams of (e), (g) and (h) are the major contribution to the EM form 
factor.  This tendency was already observed in Ref.~\cite{Ito:1991pv}
as well as in Ref.~\cite{Dorokhov:2003sc}.  However, in contrast, the
contributions from diagrams (a), (b) and (f) are relatively
small and zero at $Q^2=0$.  Moreover, we observe that the charge radii
($\langle r^2\rangle_{\cal M}$) sensitively depend on them.  In particular,
the cancellation between the diagrams (b) and (f) plays an
important role in producing the value of the pion EM charge radius
compatible to the experimental one.  

\begin{figure}[t]
\begin{tabular}{cc}
\includegraphics[width=7.7cm]{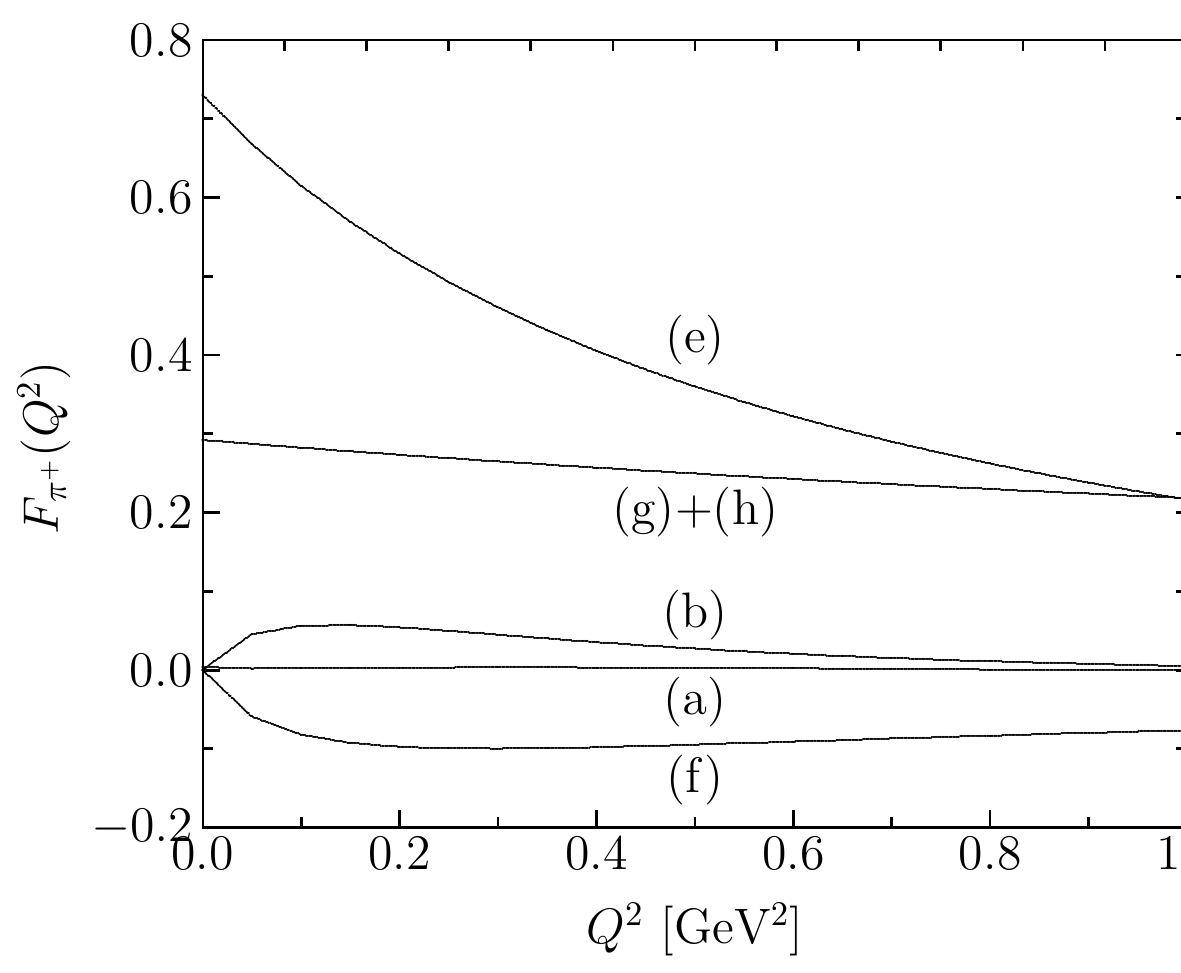}
\includegraphics[width=7.5cm]{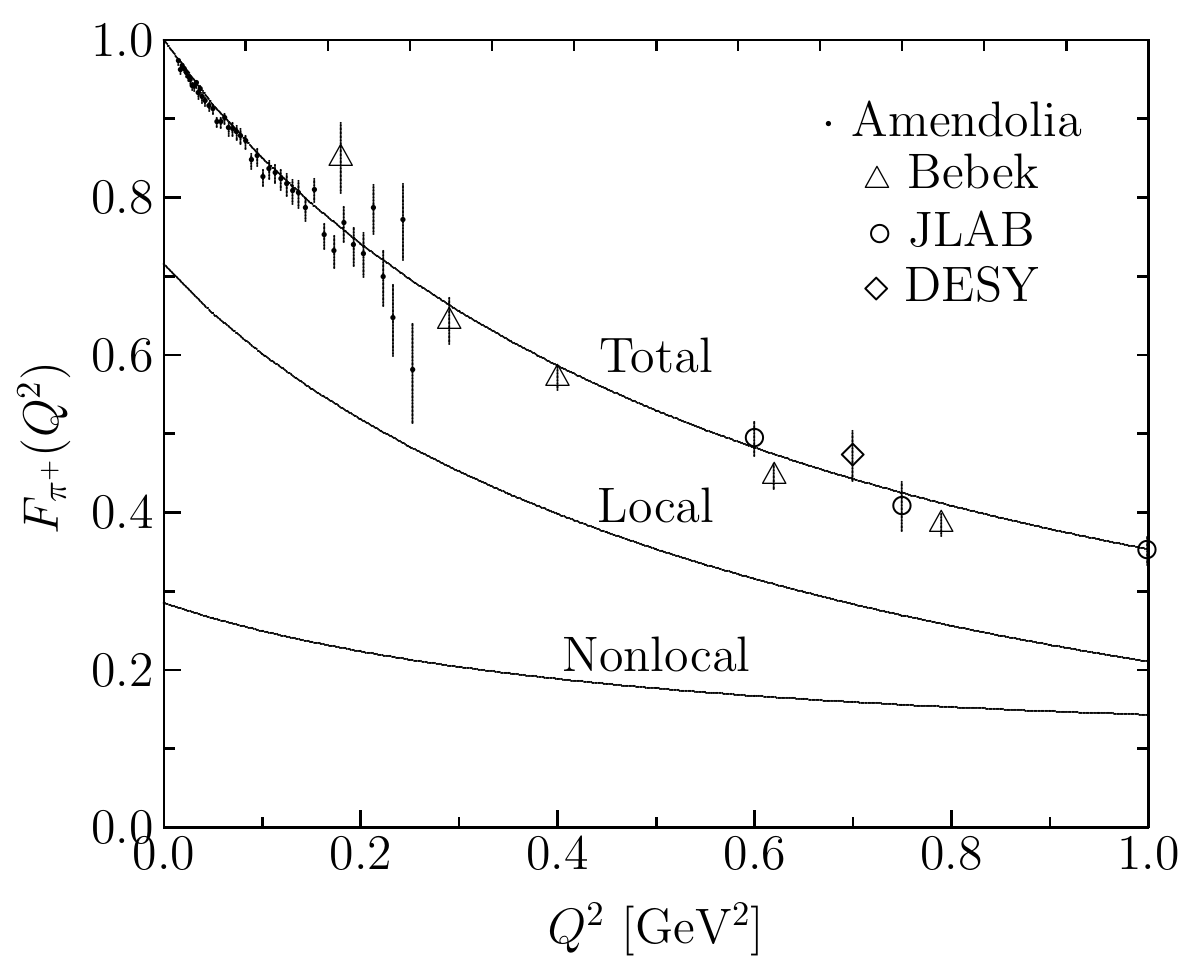}
\end{tabular}
\begin{tabular}{cc}
\includegraphics[width=7.5cm]{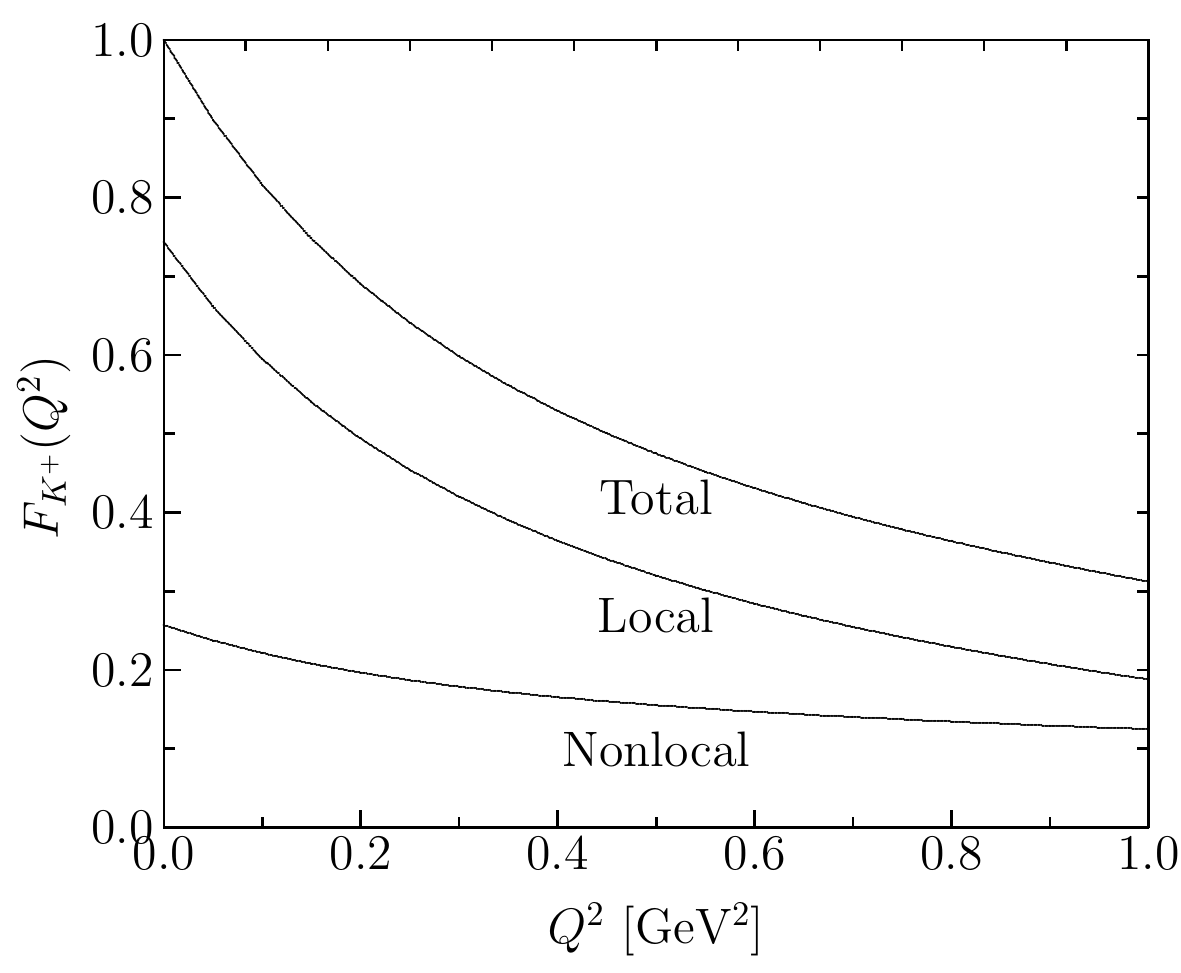}
\includegraphics[width=7.5cm]{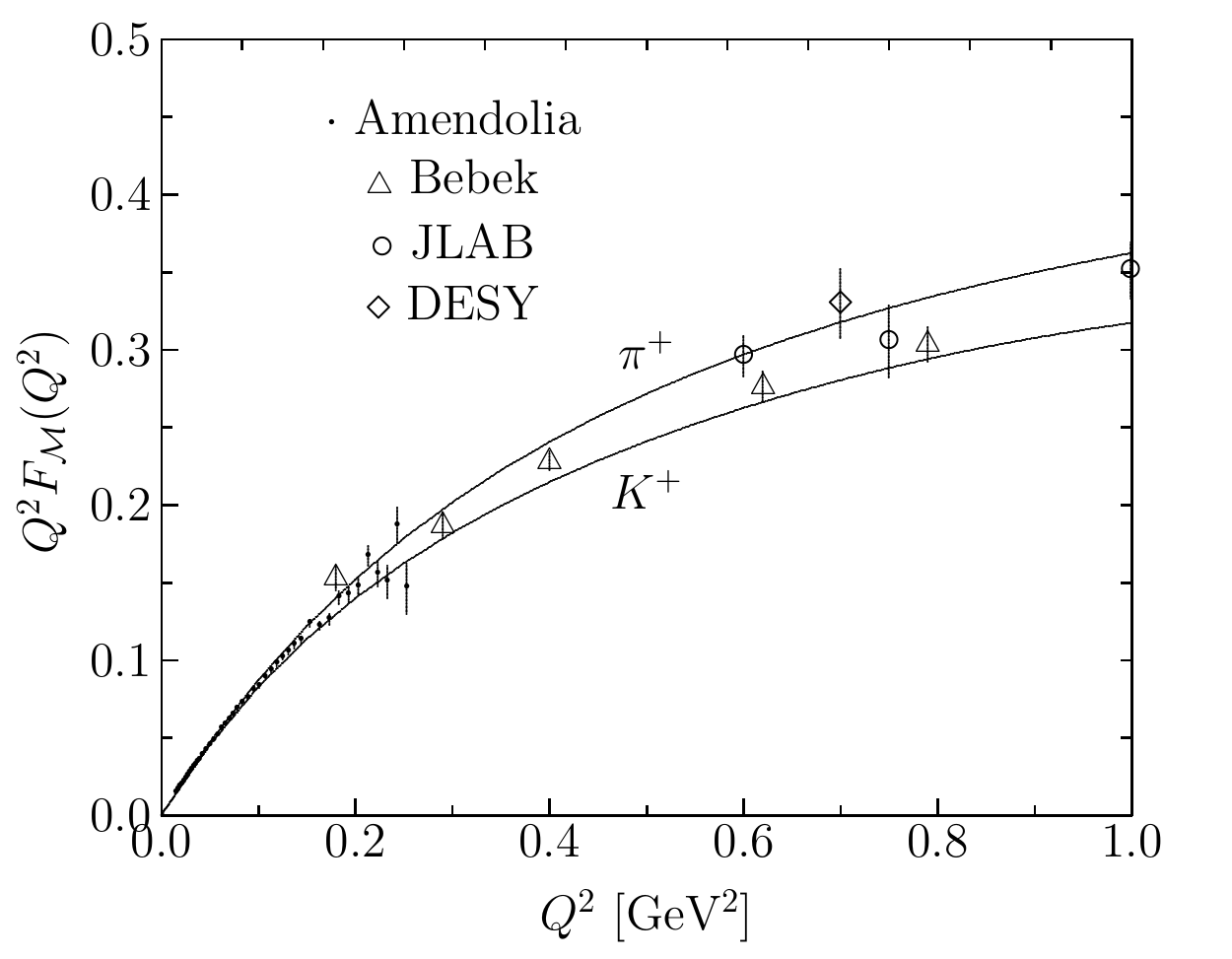}
\end{tabular}
\caption{Contribution from each diagram, shown in Fig.~\ref{fig:1}
for $F_{\pi^+}(Q^2)$ (upper left). $F_{\pi^+}(Q^2)$ (upper right)
and $F_{K^+}(Q^2)$ (lower left) depicted with the local and nonlocal
contributions separately. $Q^2F_{K^+}(Q^2)$ and $Q^2F_{K^+}(Q^2)$
are also given (lower right).  The experimental data for
$F_{\pi^+}(Q^2)$ are taken from Refs.~\cite{Bebek:1974iz,Bebek:1974ww}
($\triangle$), \cite{Amendolia:1986wj} ($\cdot$),
\cite{Volmer:2000ek} ($\circ$) and \cite{Brauel:1979zk} ($\diamond$).}        
\label{fig:2}
\end{figure}

The results of the pion and kaon EM form factors $F_{\pi^+}(Q^2)$ and
$F_{K^+}(Q^2)$ are drawn in the upper-right and lower-left panels,
respectively, in Fig.~\ref{fig:2}.  The $F_{K^0}$ is not presented,
since it is almost zero in the region: ($0\lesssim Q^2\lesssim1$ 
GeV).  The experimental data for $F_{\pi^+}(Q^2)$ are taken from
Refs.~\cite{Bebek:1974iz,Bebek:1974ww,Amendolia:1986wj,Volmer:2000ek}.
We test the normalization conditions for the form factors given  
in Eq.~(\ref{eq:norm}).  Using the model values for the pion and kaon
weak decay constants, $F_{\pi}=93$ MeV and $F_K=108$
MeV~\cite{Nam:2007fx}, which are almost the same with the empirical
values, we have $F_{\pi^+}(0)=F_{K^+}(0)=1$ and
$F_{K^0}(0)=0$ without adjusting any parameter.  It indicates that the
current conservation is preserved in the present work.  Here, we
want to discuss the charge normalization condition for the form factor
in detail. For definiteness, we consider diagrams (e), (g) and (h) 
which survive at $Q^2=0$ as shown in the upper-left panel of
Fig.~\ref{fig:2}.  Taking the limit $Q^2\to 0$, we have $k_a\to k-P/2$
and $k_{b,c}\to k+P/2$ in which $P=P_i=P_f$~\cite{Nam:2007fx}. Using
them, the contributions from diagrams (e), (g) and (h) in
Eq.~(\ref{eq:EMFF}), the pion EM form factor at $Q^2=0$ can be
rewritten in the chiral limit as follows:   
\begin{eqnarray}
\label{eq:norm1}
\lim_{q\to0}F_{\pi}(Q^2)&\approx&
\frac{4N_c}{F^2_{\pi}}
\int^{\infty}_{-\infty}\frac{d^4k}{(2\pi)^4}
\left[\frac{M^2}{(k^2+M^2)^2
}+\frac{2M\sqrt{M}\tilde{M}}{k^2+M^2} 
\right]+{\cal O}(k\cdot P,m^4_{\pi}),
\end{eqnarray}
where $M'$ is defined as
\begin{equation}
\label{eq:mdash}
\tilde{M}=-\frac{1}{|k|}\frac{\partial\sqrt{M}}{\partial
  |k|}=\frac{4\sqrt{M_0}\Lambda^2}{(2\Lambda^2+k^2)^2}. 
\end{equation}
In deriving Eq.~(\ref{eq:norm1}), we have assumed $P\ll1$ for
simplicity.  Since the pion decay constant $F_{\pi}$ is expressed in
the present model as follows: 
\begin{equation}
\label{eq:FPI}
F^2_{\pi}\approx4N_c
\int^{\infty}_{-\infty}\frac{d^4k}{(2\pi)^4}
\left[\frac{M^2}{(k^2+M^2)^2
}+\frac{2M\sqrt{M}\tilde{M}}{k^2+M^2} 
\right]+{\cal O}(k\cdot P,m^4_{\cal M}),
\end{equation}
we show that the pion form factor calculated in the present model
satisfies correctly the following normalization condition:
\begin{equation}
\lim_{q\to0}F_{\pi}(Q^2) = 1.  
\end{equation}
We again want to emphasize that the normalization condition of
Eq.~(\ref{eq:norm}) is satisfied without any adjustment of
parameters.  In other words, once we choose the $1/\bar{\rho}\simeq
600$ MeV (equivalently $\Lambda$), then the $M_0$ and $F_{\pi}$ are
uniquely determined within the model (by the saddle-point equation of 
Eq.~(\ref{eq:saddle}) for instance).

We have verified that the charge normalization condition beyond
the chiral limit ($F_K(0)$) is also satisfied by  
Eqs.~(\ref{eq:norm1}) and (\ref{eq:FPI}), replacing $M$ by $m+M$ and
with the model value for the kaon decay constant $F_K$. However, we
note that the value of the $F_K$ in the present work, compared to the
empirical value $F_K=113$ MeV, turns out to be underestimated:
$F_K\approx108$ MeV. This can be understood, since we take the large
$N_c$ limit.  The meson-loop corrections ($1/N_c$ corrections) are
known to be essential in improving the value of the $F_K$.  Thus,
within the present framework, the charge normalization conditions are
satisfied consistently for the kaon EM form factor as well as the pion 
one.  

We draw the local and nonlocal contributions separately, in addition
to the total one in the lower left panel of Fig.~\ref{fig:2}.  We see
also from Fig.~\ref{fig:2} that the $F_{\pi^+}(Q^2)$ is reproduced
quantitatively well in comparison to the experimental data. It turns
out that the nonlocal terms contribute to the form factors by about
$30\,\%$.  Thus, it is of great importance to consider the gauge 
invariance from the outset.  Note that the overall shape and behavior
of the kaon EM form factor $F_{K^+}$ are rather similar to
$F_{\pi^+}(Q^2)$ apart from the fact that it falls off slightly 
faster than that of the pion.  In the lower right panel of
Figure~\ref{fig:2}, we plot $Q^2F_{\pi^+,K^+}(Q^2)$.
\begin{table}[b]
\begin{tabular}{c||c|c|c|c}
&Local&Nonlocal&Total&Exp.~\cite{PDG}\\
\hline
$\langle r^2\rangle_{\pi^+}^{1/2}\,[\mathrm{fm}] $& $0.594$ & $0.319$
& $0.675$ & $0.672\pm 0.008$\\ \hline
$\langle r^2\rangle_{K+}^{1/2}\,[\mathrm{fm}] $&$0.658$ & $0.318$ &
$0.731$ & $0.560\pm 0.031$ \\ 
\hline
$\langle r^2\rangle_{K^0}\,[\mathrm{fm}^2]
$&$-0.044$&$-0.016$&$-0.060$& $-0.077\pm 0.010$\\ 
\end{tabular}
\caption{Each contribution to the pion and kaon EM charge radii.  We 
present also the mean square charge radius for the neutral kaon 
($K^0$).  The abbreviation Exp. denotes the experimental data taken
from Ref.~~\cite{PDG}.} 
\label{table1}
\end{table}

We now examine the pion and kaon charge radii $\langle
r^2\rangle_{\cal M}^{1/2}$, defined as Eq.~(\ref{eq:radi}).  In 
Table~\ref{table1} we list the results of the pion and kaon charge
radii ($\pi^+$, $K^+$ and $K^0$) with each contribution
separately.  For the neutral kaon ($K^0$), we list the mean 
square charge radius.  Note that these values are well compatible with 
those computed from the general QCD parameterization
method~\cite{Dillon:2001si}.  As seen in the case of the EM form
factors, we find that the nonlocal part contributes to the radii by
about $30\%$ again.  The nonlocal contributions obtained in the
present work are more significant, compared with those of 
Refs.~\cite{Ito:1991pv,Dorokhov:2003sc}, in which they
are estimated to be only $\sim10\%$.  On the contrary, the local
contribution is rather similar to that of Ref.~\cite{Dorokhov:2003sc}
($0.55\,{\rm fm}$).  Interestingly, we obtain the $K^+$ charge radius
about $30\%$ lager than the experimental data.  As mentioned
previously, the $1/N_c$ meson-loop corrections are known to be
essential in describing the kaon EM form factors, which we have neglected
in the present work.  Actually, we have drawn a similar conclusion in
the case of the kaon semileptonic form factors~\cite{Nam:2007fx}. The
mean square charge radius of the neutral kaon $\langle
r^2\rangle_{K^0}$ is also slightly  underestimated in the present work
by about $10\%$, which is again due to the absence of the 
$1/N_c$ meson-loop corrections.   

In Ref.~\cite{Roberts:1994hh}, nonperturbative Dyson-Schwinger (DS)
and Bether-Salpeter (BS) equations were employed with the confining
quark propagator and dressed quark-photon vertex similar to our
framework.  However, being different from ours, the nonlocal
interaction has been taken into account explicitly in
Ref.~\cite{Roberts:1994hh} in addition to the local ones, and
parameters for the algebraic quark propagator were fitted with the
LECs in the local impulse approximation.
Refs.~\cite{Ito:1991pv,Buck:1994hf} used the generalized  
NJL model in which the separable interaction was considered for the 
pion and kaon charge radii.  Their values were well
comparable with the experimental data as well as with $\chi$PT
results~\cite{Bijnens:2002hp}.  In those works, there were two
parameters, namely, constituent quark mass at $q^2=0$ and the cutoff
mass $\Lambda$, which were fitted to LECs, $F_{\pi,K}$ and its
two-photon decay width.  Note that in the present work the parameter
$M_0$ is fixed by the saddle-point equation and the strange current
quark mass $m_{\rm s}$ is fitted to the kaon mass. Thus, the model is
much constrained from the instanton vacuum.  

In $\chi$PT, the pion EM form factor has been already studied to 
order $\mathcal{O}(p^6)$~\cite{Bijnens:2002hp}.  The pion EM charge 
radius is quite comparable with our results, whereas that
for the $K^+$ is about $30\%$ smaller than the present one.  This
discrepancy tells again that the meson-loop corrections are essential
in describing the kaon EM form factor.  

Finally, we discuss the Gasser-Leutwyler LEC $L_9$ in the large $N_c$
limit, which can be determined by the pion EM charge radius.  From the
present numerical value for $\langle r^2\rangle_{\pi^+}$, we obtain
$L_9=8.42\times10^{-3}$.  In $\chi$PT~\cite{Bijnens:2002hp}, the
renormalized LEC $L_9^r$ at $\mu=0.77\,\mathrm{GeV}$ has been
determined by using the data for the pion and kaon EM form factors:
$(5.93\pm 0.43)\times 10^{-3}$.  This difference turns out to be
around $30\,\%$.
  
\section{Summary and conclusion}
In the present work, we have investigated the pion and kaon
electromagnetic form factors in the range of $0\le Q^2\le1$
GeV from the instanton vacuum with explicit SU(3) symmetry
breaking effects taken into account.  We have briefly shown how to
construct the gauged effective chiral action following
Refs.~\cite{Musakhanov:2002xa,Kim:2004hd}.  There are eight
independent Feynman diagrams (local and nonlocal, isosinglet and
isotriplet) for the pion and kaon electromagnetic form factors.
Among them, two diagrams including a single quark propagator vanish.
It turned out that the nonlocal corrections contribute to the pion EM
form factors by about $30\%$.  We stress that all relevant diagrams
for the pion and kaon EM form factors were generated without any
phenomenological assumptions, the conservation of the electromagnetic
current being satisfied.  Moreover, all parameters, {\it i.e.} the
constituent quark mass at the zero virtuality and the current quark
mass are already fixed from the instanton vacuum, so that we do not
have any additional free parameter to adjust.  These features provide
considerable merit to describe the pion and kaon electromagnetic form
factors.    

First, we have computed the pion and kaon form factors
($F_{\pi^+,K^+,K^0}(Q^2)$).  We have observed that the numerical
results for the pion electromagnetic form factor $F_{\pi^+}(Q^2)$ are
in good agreement with the experimental data.  While the
$F_{K^+}(Q^2)$ turned out to be similar to that of the pion in its 
overall shape, it falls off faster than the pion one as $Q^2$
increases.  The $F_{K^0}(Q^2)$ is almost negligible for the considered
region in comparison to those of $\pi^+$ and $K^+$: 
$|F_{K^0}(Q^2)|\lesssim10^{-2}$.   

The pion and kaon EM charge radii have been calculated as well.  The
numerical results are summarized as follows: 
$\langle{r^2}\rangle^{1/2} _{\pi^+} = 0.675\,\mathrm{fm}$, 
$\langle{r^2}\rangle^{1/2} _{K^+} = 0.731\,\mathrm{fm}$ and  
$\langle r^2\rangle_{K^0}=-0.060\,\mathrm{fm}^2$ without any
adjustable free parameters again.  Compared to the experimental data,
the result of the pion electromagnetic charge radius is in very good
agreement with the data, while that of the $\langle r^2\rangle_{K^+}$
is about $30\,\%$ larger than the experimental one, which indicates
that the $1/N_c$ meson-loop corrections are required to describe the
kaon properties quantitatively~\cite{Nam:2007fx}.  From the 
pion charge radius, we were able to extract the low-energy constant
$L_9=8.42\times10^{-3}$.  

In conclusion, the nonlocal chiral quark model from the instanton
vacuum with SU(3) symmetry breaking effects provides a good framework
to study the pion and kaon EM form factors.  Moreover, all the
relevant contributions to the EM form factors were generated in a
simple and systematic way.  However, in order to investigate kaonic
properties within the present framework, the $1/N_c$ meson-loop
contributions~\cite{Goeke:2007bj,Goeke:2007nc} are essential for which
we consider as our future work.  

\section*{Acknowledgments}
The authors are grateful to fruitful discussions with K. Goeke,
A.~Hosaka, M. Siddikov, M.~M.~Musakhanov and A.~Dorokhov. S.i.N. is
thankful to T.~Kunihiro and Y.~Kwon for valuable discussions.  The
present work is supported by the Korea Research Foundation Grant
funded by the Korean Government(MOEHRD) (KRF-2006-312-C00507).  The
work of S.i.N. is partially supported by the grant for Scientific
Research (Priority Area No.~17070002) from the Ministry of Education,
Culture, Science and Technology, Japan.  The numerical calculations
were carried out on YISUN at YITP in Kyoto University and on MIHO at
RCNP in Osaka University.  
 
\section*{Appendix}
\noindent
Definition of the momenta in the Breit-momentum frame:
\begin{eqnarray}
  \label{eq:k}
p_i&=&\left(0,\,\,\,0,\,\,\,\frac{Q}{2},\,\,\,i\sqrt{m^2_{\cal{M}}
+\frac{Q^2}{4}}\right),\,\,\,\,q=\left(0,\,\,\,0,\,\,\,iQ,\,\,\,0\right),
\nonumber\\
k&=&\left(k_r\sin\phi\sin\psi\cos\theta,\,\,\,
k_r\sin\phi\sin\psi\sin\theta,\,\,\,
k_r\sin\phi\cos\psi,\,\,\,k_r\cos\phi\right),
\nonumber
\end{eqnarray}
where the angles ($\phi$, $\psi$, $\theta$) are defined in the
four-dimensional Euclidean integral as follows: 
\begin{eqnarray}
  \label{eq:integral}
\int^{\infty}_{-\infty}{d^4k}=\int_0^{\infty}k^3_rdk_r
\int_{0}^{\pi}\sin^2\phi\,d\phi
\int^{\pi}_{0}\sin\psi\,d\psi\int_0^{2\pi}d\theta.     
\nonumber
\end{eqnarray}

\end{document}